\documentclass[
twocolumn,
amsmath,
amssymb,
aps,
prl,
nofootinbib,
superscriptaddress, 
floatfix,
longbibliography]{revtex4-2}

\usepackage{newtxtext, newtxmath}

\usepackage[pdftex]{graphicx}
\usepackage{hyperref}
\usepackage{physics}
\usepackage{bm}
\usepackage{braket}
\usepackage{amsfonts,mathrsfs}
\usepackage{mathtools}
\usepackage{xspace}
\usepackage{array}
\usepackage{multirow}
\usepackage{comment}

\hypersetup{colorlinks = true, urlcolor = blue, linkcolor = blue, citecolor = blue}

\makeatletter

\let\MYcaption\@makecaption
\makeatother

\usepackage{subcaption}
\makeatletter
\let\@makecaption\MYcaption
\makeatother

\usepackage{color}
\usepackage[normalem]{ulem} 

\newcommand{\ul}{\underline}

\begin{document}

\title{Superconducting diode effect in correlated electron systems by nonreciprocal magnetism}

\author{Kyohei Nakamura}
\email[]{nakamura.kyohei.84w@st.kyoto-u.ac.jp}
\affiliation{Department of Physics, Graduate School of Science, Kyoto University, Kyoto 606-8502, Japan}

\author{Youichi Yanase}
\affiliation{Department of Physics, Graduate School of Science, Kyoto University, Kyoto 606-8502, Japan}

\date{\today}

\begin{abstract}
The superconducting diode effect (SDE), characterized by a nonreciprocal critical current in superconductors, has recently been observed in strongly correlated electron systems and near quantum criticality, pointing to unconventional mechanisms beyond weak-coupling theories. 
Here we investigate the SDE in the Rashba-Zeeman-Hubbard model, which captures $d$-wave superconductivity in an antiferromagnetic quantum critical regime, using the Dyson-Gor'kov equation with the fluctuation exchange approximation.
We show that electron correlations suppress the conventional intrinsic SDE arising from depairing currents.
More importantly,
a supercurrent nonreciprocally induces antiferromagnetic order, which fundamentally governs the critical current and enables perfect diode efficiency.
Our results reveal a previously unrecognized correlation-driven mechanism of the SDE and establish strongly correlated superconductors as a platform for superconducting diode physics.
\end{abstract}

\maketitle

\textit{Introduction.}---A critical current in bulk superconductors lacking inversion and time-reversal symmetries can be nonreciprocal, such that the system switches between superconducting and resistive states depending on the current direction. 
This nonreciprocal critical current is called the superconducting diode effect (SDE)~\cite{Ando2020,Nadeem2023-jf,Nagaosa2024}. 
Because of its potential for dissipation-free diode functionality and for probing of symmetry breaking, the SDE has attracted considerable interest. Theoretically, it has been studied in terms of intrinsic SDE based on depairing currents~\cite{Yuan2022,Daido2022,Ilic2022-it,He_2022,Daido2022prb}, extrinsic SDE arising from vortex motion~\cite{Hou2023-if}, and even the perfect SDE~\cite{Daido2025-uni,Shaffer2024-ql,Chakraborty2025-aw}. 
Experimentally, it has been reported in Rashba heterostructures~\cite{kawarazaki2022-ux,Narita2022}, transition metal dichalcogenides~\cite{Bauriedl2022,He2025-wv}, twisted multilayer graphene~\cite{Lin2022}, and, more recently, in strongly correlated electron systems (SCES) such as high-$T_{\rm c}$ cuprates~\cite{Mizuno2022,Qi2025}, $\rm{Fe}$-based superconductors~\cite{Nagata2025-nc,Dong2025-ag,Kobayashi2025-hx}, and kagome superconductors~\cite{Le2024}, some of which lie near quantum criticality.

Despite considerable interest in unconventional superconductivity in SCES, the microscopic mechanism of the SDE in SCES remains poorly understood because most previous studies have been based on mean-field approaches that do not capture quantum fluctuations and criticality characteristic of SCES. 
In this work, we show that strong correlations qualitatively alter the mechanism of the SDE: a supercurrent nonreciprocally induces antiferromagnetic (AF) order, leading to a fundamentally distinct, correlation-driven SDE.

Establishing the SDE and supercurrent-induced phenomena in SCES is important for three reasons. 
First, SCES can exhibit high superconducting transition temperatures and large critical currents, which are advantageous for device applications~\cite{Brooks2025-vk,Wang2025-pn}. 
Second, the supercurrent can serve as a novel tuning parameter of quantum criticality~\cite{Nakamura2025}. 
Third, the SDE in SCES may originate from unconventional mechanisms arising from strong correlations and quantum fluctuations that also drive unconventional superconductivity.

To clarify the SDE in quantum-critical $d$-wave superconductors near the AF instability, as typified by high-$T_{\rm c}$ superconductors, we study a two-dimensional Rashba-Zeeman-Hubbard model. 
Furthermore, to incorporate strong correlations and AF critical fluctuations, we employ the Dyson-Gor'kov equation with the fluctuation exchange (FLEX) approximation~\cite{Moriya01072000,YANASE20031,Bickers1989,BICKERS1989206,Kita2011-at}, which captures momentum-dependent spin fluctuations beyond local approximations. 
Compared with dynamical mean-field theory, which lacks nonlocal correlations~\cite{Gerorges1996}, and quantum Monte Carlo methods, which may suffer from the sign problem and limited momentum resolution in finite-size systems~\cite{Troyer2005}, our approach is well suited to the present study.

Our main results are twofold. 
First, the strong electron correlations suppress the intrinsic SDE determined by depairing currents. 
Second, the supercurrent nonreciprocally induces AF order, and this nonreciprocal magnetism leads to a correlation-driven mechanism of the SDE. 
In particular, it can yield a perfect SDE with the diode efficiency of $100[\%]$. 
These results suggest that the supercurrent serves as a control parameter that tunes the interplay of magnetism and unconventional superconductivity, thereby realizing unique functionalities of strongly correlated superconductors.

\textit{Formulation.}---We start from the two-dimensional Rashba-Zeeman-Hubbard model.
The Hubbard model describes AF quantum criticality and $d$-wave superconductivity.
The Rashba spin-orbit coupling and the Zeeman effect are introduced because the SDE requires the breaking of inversion and time-reversal symmetries.
Many previous theories of the SDE~\cite{Yuan2022,Daido2022,Ilic2022-it,Daido2022prb} adopted the Rashba-Zeeman model, but the effects of electron interactions and quantum criticality are neglected there.
The Hamiltonian including the Hubbard interaction is given by
\begin{align}
    H=&\sum_{\bm{k}\sigma\sigma'} \left\{ \varepsilon(\bm{k})\delta_{\sigma\sigma'}+[\bm{g}(\bm{k})-\bm{h}]\cdot\bm{\sigma}_{\sigma\sigma'} \right\} c^\dagger_{\bm{k}\sigma}c_{\bm{k}\sigma'} \notag \\
    &+U\sum_{i}n_{i\uparrow}n_{i\downarrow},
\end{align}
where $\bm{k}$, $\sigma=\uparrow,\downarrow$, and $i$ are indices of momentum, spin, and site, respectively, $c^\dagger_{\bm{k}\sigma}$ ($c_{\bm{k}\sigma}$) is an electron creation (annihilation) operator, 
and $n_{i\sigma}$ is an electron density operator at the site $i$ with spin $\sigma$.

We consider a square lattice and a tight-binding energy dispersion,
$
    \varepsilon(\bm{k})=-2t(\cos k_x +\cos k_y)+4t'\cos k_x \cos k_y -\mu
$.
The chemical potential $\mu$ is chosen so that the density of electrons is $n=0.85$, which corresponds to the optimally doped region of high-$T_{\rm c}$ cuprate superconductors. 
The Rashba type $g$ vector is given by
$
   \bm{g}(\bm{k})=\alpha\qty[-\partial\varepsilon(\bm{k})/\partial{k_y},\partial\varepsilon(\bm{k})/\partial{k_x},0]
$ \cite{Yanase2007-soc,Yanase2008-soc}.
The Zeeman field is applied in the $y$ direction, $\bm{h}=(0,h,0)$.
We set $t'=0.3$ and $\alpha=0.3$ with a unit of energy $t=1$.
To study the supercurrent-carrying state and the SDE, we assume Cooper pairs with finite momentum $2\bm{p}$ in the $x$ direction.
For later use, we introduce a $2\times2$ matrix for single-particle Hamiltonian given by
$
    \ul{\xi}(\bm{k})=\varepsilon(\bm{k})\bm{I}_2+[\bm{g}(\bm{k})-\bm{h}]\cdot\bm{\sigma}
$,
where $\bm{I}_n$ and $\bm{\sigma}$ are the $n\times n$ identity matrix and the Pauli matrix, respectively.
In the following,
$\ul{A}$, $\hat{A}$, and $\check{A}$ denote $2\times2$, $4\times4$, and $8\times8$ matrices, respectively.

The formulation of the Dyson–Gor'kov equations and the FLEX approximation is given in the Supplemental Material~\cite{supplemental}.
We adopt the FLEX approximation for two reasons.
First, it incorporates critical spin fluctuations and thus captures the essential properties of the AF quantum critical $d$-wave superconductivity~\cite{Moriya01072000,YANASE20031}.
Second, it is a conserving approximation~\cite{Baym1961} that satisfies thermodynamic relations, and thus we can uniquely evaluate thermodynamic quantities such as the supercurrent, which is necessary to investigate the SDE, from a thermodynamic potential~\cite{supplemental}. 
In the FLEX approximation, the order parameter of superconductivity, magnetic susceptibility, and thermodynamic quantities are self-consistently determined. The temperature is set to $T=0.005<T_{\rm c} \sim0.015$, which is much lower than the superconducting critical temperature $T_{\rm c}$.

Now we introduce key physical quantities such as the Stoner factor, supercurrent, and superconducting diode efficiency. 
First, the Stoner factor is defined as follows.
The $8\times8$ matrix of spin susceptibility is calculated by
$
    \check{\chi}(q,\bm{p})=\check{\chi}^0(q,\bm{p})\qty[\bm{I}_8+\check{U}\check{\chi}^0(q,\bm{p})]^{-1} 
$ with the bare interaction matrix $\check{U}$ and the bare spin susceptibility matrix $\check{\chi}^0(q,\bm{p})$ whose explicit forms are given in  \cite{supplemental}. 
Then, the Stoner factor $\alpha_S(\bm{p})$ is defined by
\begin{align}
    \alpha_{\rm S}(\bm{p})=\max_q\max_{e.v.}\qty[-\check{U}\check{\chi}^0(q,\bm{p})], \label{eq:stoner}
\end{align}
where $\max_{e.v.}[A]$ denotes the maximum eigenvalue of $A$.
Second, we formulate the supercurrent and define the superconducting diode efficiency.
The supercurrent $j(\bm{p})$ along the {\it x}-axis is obtained by
\begin{align}
    j(\bm{p})=\frac{1}{2}\sum_k\mathrm{Tr}\qty[\pdv{\hat{\Xi}(\bm{k},\bm{p})}{k_x}\hat{\mathcal{G}}(k,\bm{p})] , \label{eq:supercurrent}
\end{align}
with
\begin{align}
    \hat{\Xi}(\bm{k},\bm{p})=\mqty(\ul{\xi}(\bm{k}+\bm{p}) & \bm{0} \\ \bm{0} & -\ul{\xi}^T(-\bm{k}+\bm{p})) , \label{eq:verocity}
\end{align}
and $\hat{\mathcal{G}}(k,\bm{p})$ being the $4\times4$ matrix of the Green function~\cite{supplemental}. 
The superconducting diode efficiency $\eta$ is defined in a standard way, 
\begin{align}
    \eta[\%]=\frac{j_{{\rm c}+}-|j_{{\rm c}-}|}{j_{{\rm c}+}+|j_{{\rm c}-}|}\times100,
\end{align}
where $j_{{\rm c}\pm}$ is the critical current in the positive and negative directions.
The mechanism of the critical current is discussed in the following results.

Here, we discuss the order parameter of superconductivity.
The point group symmetry of the system 
is $C_{4v}$, and it has been shown that superconductivity of $B_1$ representation is most stable in a wide parameter range~\cite{Nogaki2020}.
Accordingly, in the zero-momentum superconducting state, the momentum dependence of the spin-singlet component $\psi(\bm{k})$ and spin-triplet component $\bm{d}(\bm{k})$ of the order parameter are roughly approximated as
$
    \psi(\bm{k})\sim \cos{k_x}-\cos{k_y}
$ and 
$
    \bm{d}(\bm{k}) \sim [\sin{k_y}, \sin{k_x}, 0]
$, respectively.
Additional components are induced in the finite-momentum superconducting state, 
although we do not touch them in detail.
In the following results, the momentum of Cooper pairs $\bm{p} = (p,0,0)$ in the $x$ direction is considered, and its functions $A(\bm{p})$ are abbreviated as $A(p)$ for simplicity.

\textit{SDE by depairing mechanism.}---In this section, we investigate the SDE derived from the depairing currents~\cite{Yuan2022,Daido2022,Ilic2022-it,He_2022,Daido2022prb}.
Here, the critical current $j_{\rm{c\pm}}$ is defined as 
$
    j_{\rm{c+}}\equiv j_{\rm{d+}}=\max_p\qty[j(p)]
$ and $
    j_{\rm{c-}}\equiv j_{\rm{d-}}=\min_p\qty[j(p)]
$,
where $j_{\rm{d\pm}}$ is the depairing critical current in the positive and negative directions.

\begin{figure}[tbp]
    \centering
    \begin{tabular}{ll}
    (a)   &(b) $U=3.5$\\ 
     \includegraphics[keepaspectratio, scale=0.0325]{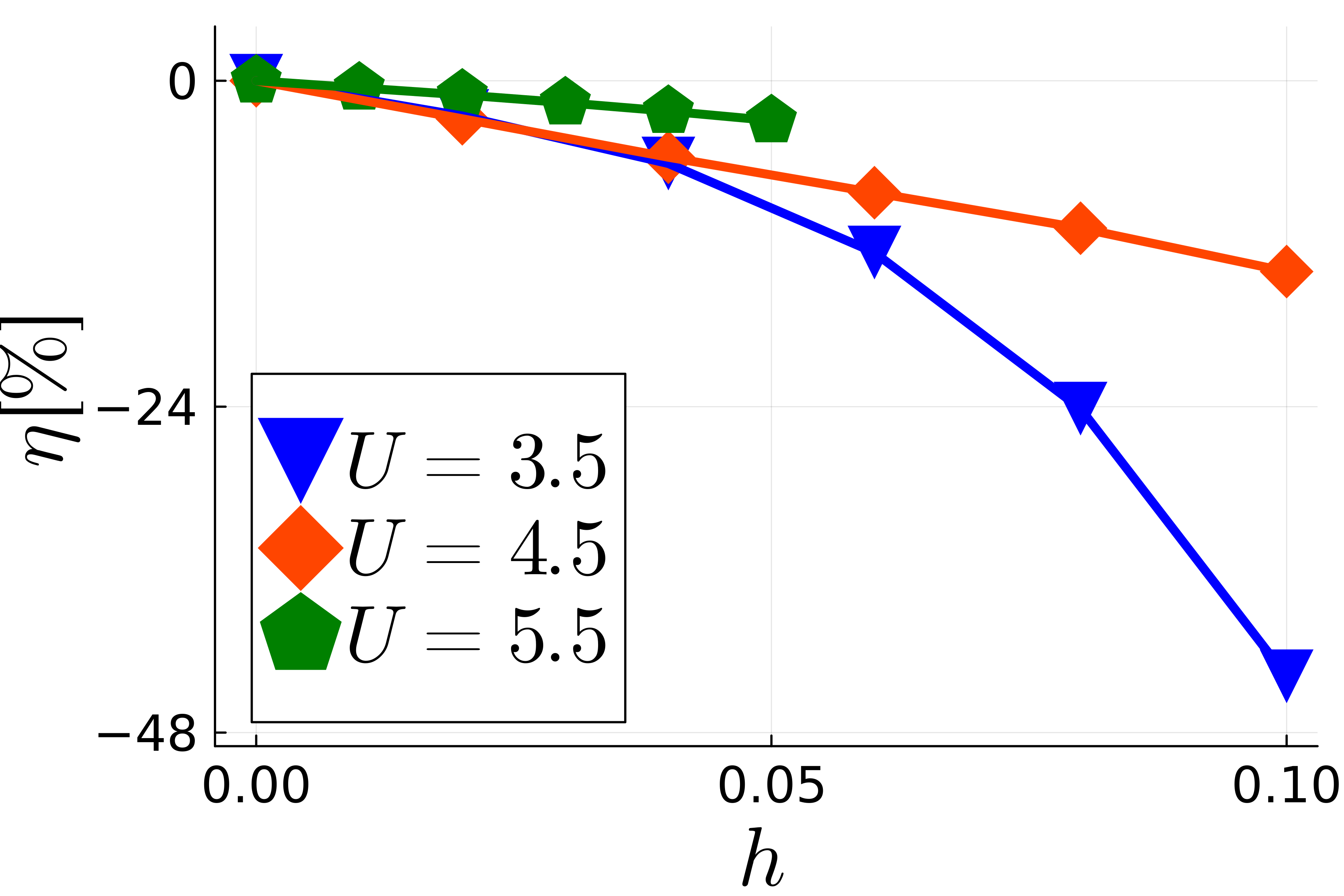} 
    &\includegraphics[keepaspectratio, scale=0.0325]{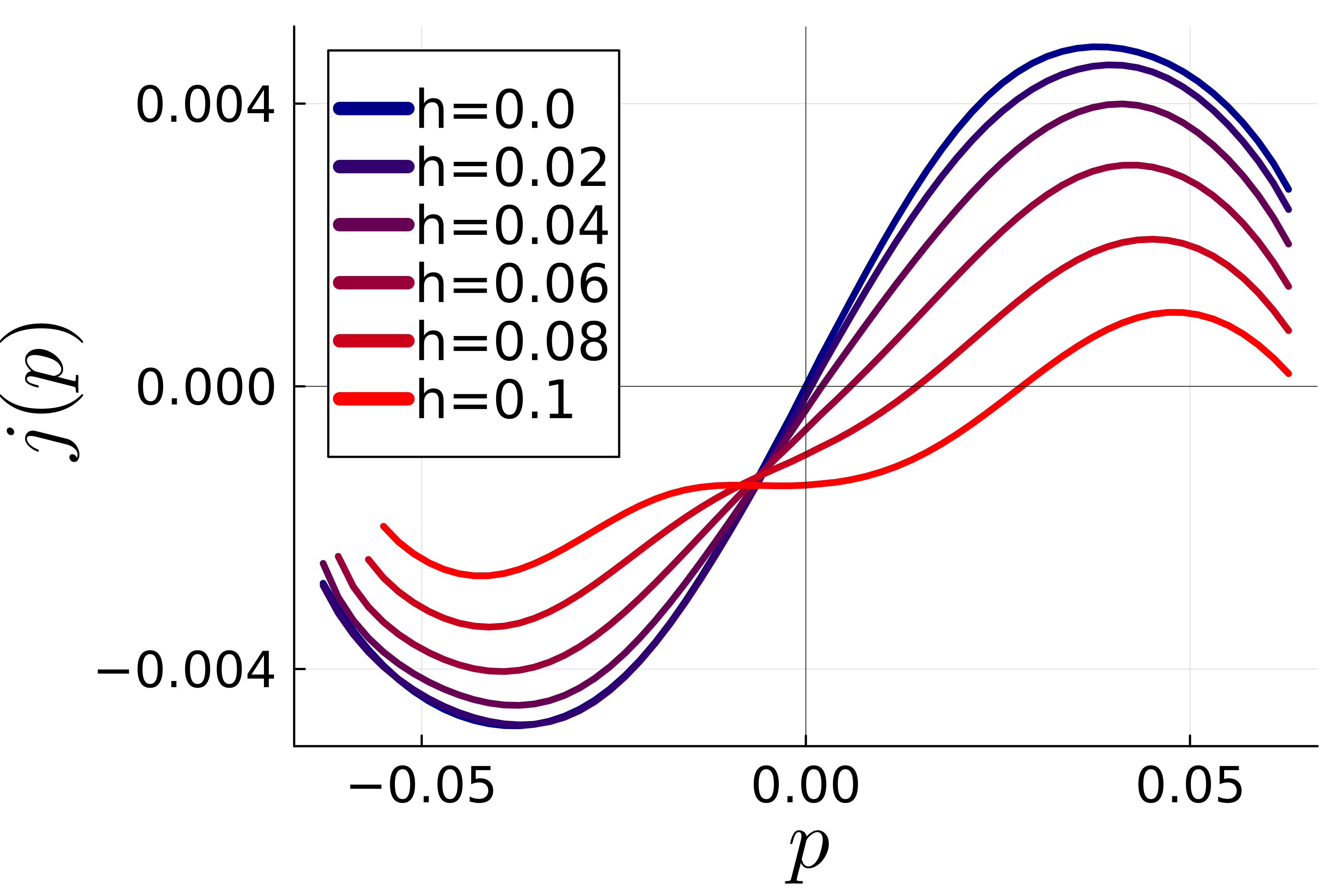} \\
    (c) $U=4.5$  &(d) $U=5.5$ \\
     \includegraphics[keepaspectratio, scale=0.0325]{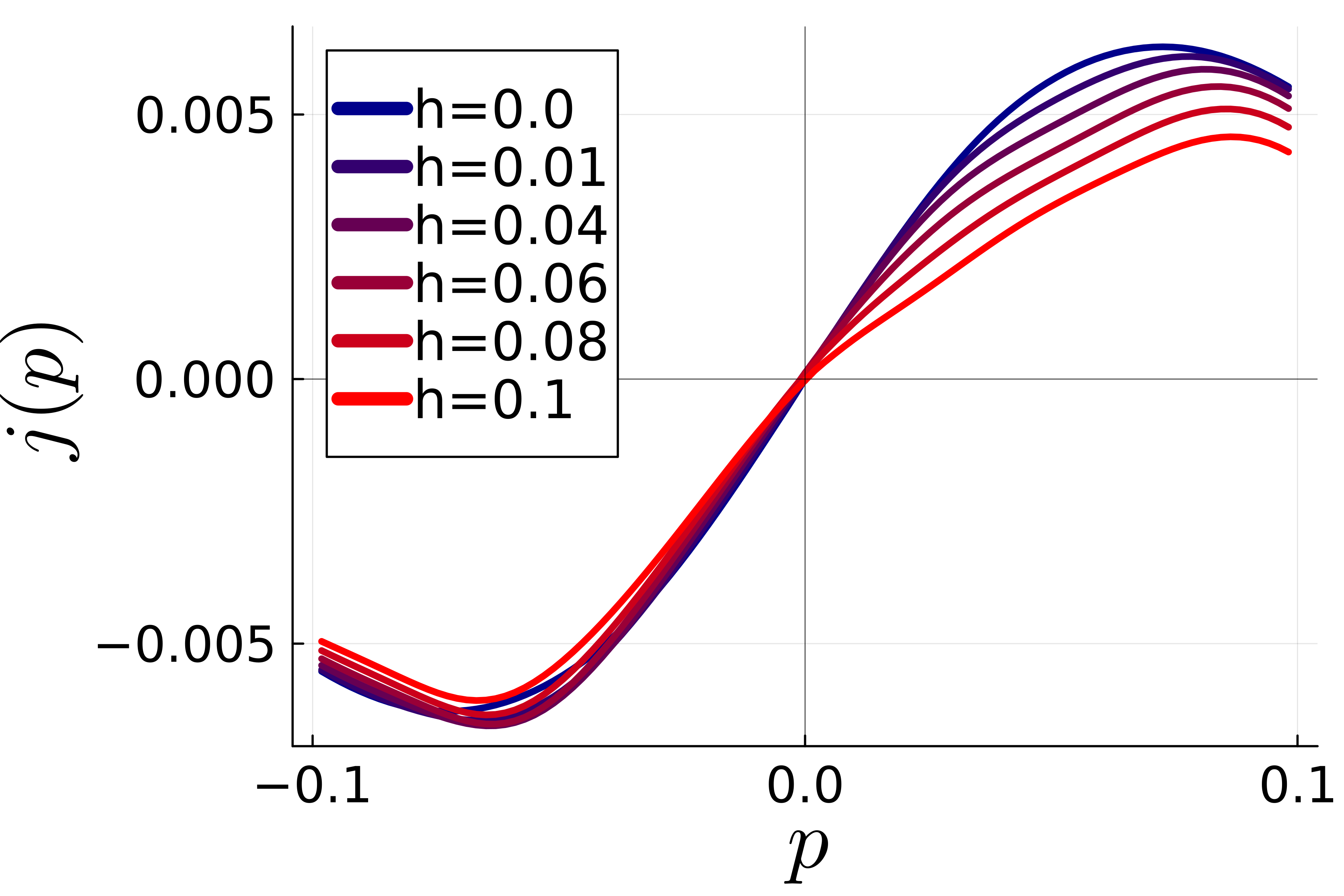} 
    &\includegraphics[keepaspectratio, scale=0.0325]{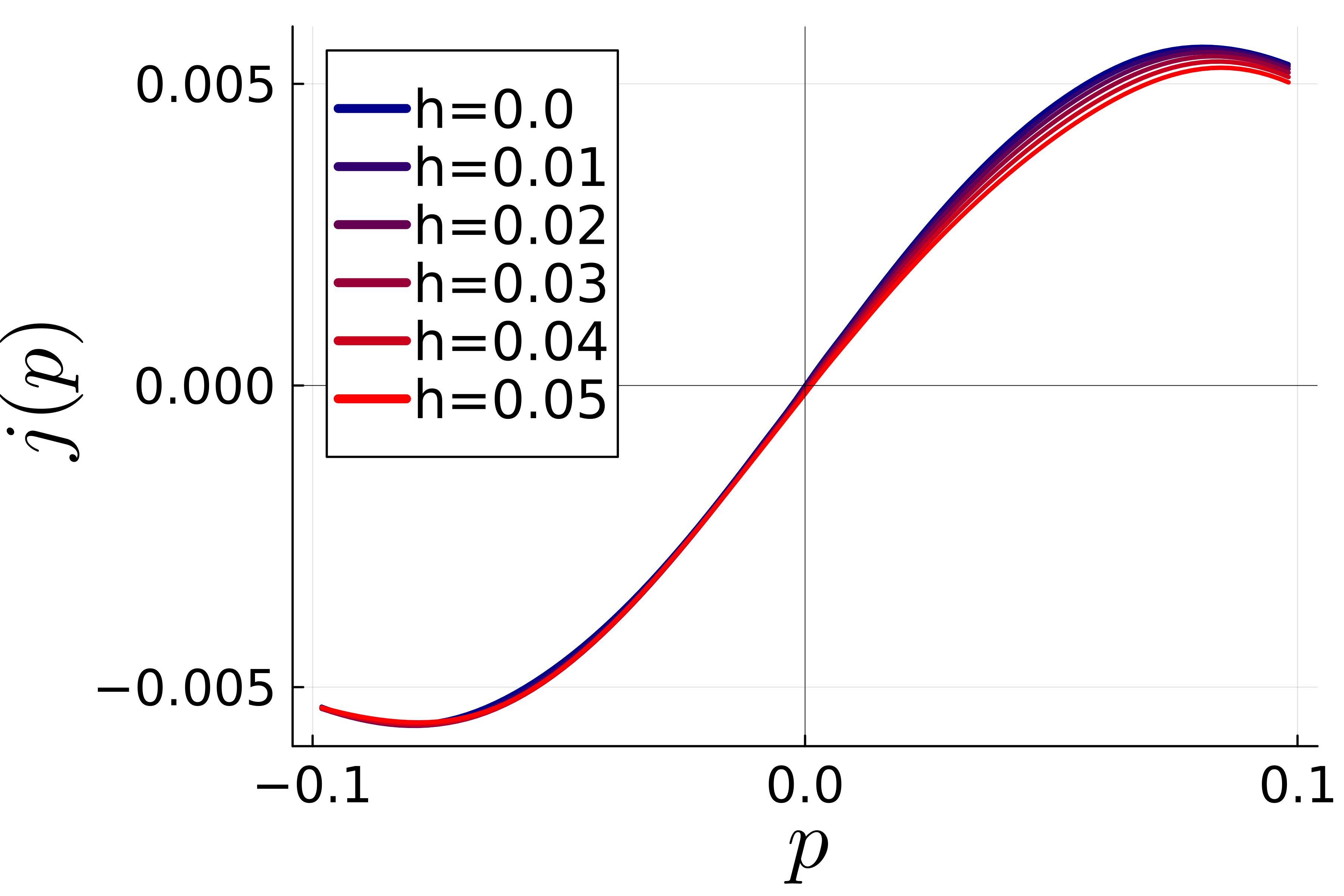}  
    \end{tabular}
    \caption{
    (a) Magnetic field dependence of the superconducting diode efficiency $\eta$ for $U=3.5$, $4.5$, and $5.5$ with triangles, diamonds, and pentagons, respectively.
    (b)-(d) Cooper pair's momentum dependence of the supercurrent $j(p)$ for various magnetic fields.
    The onsite Coulomb interaction is varied as (b) $U=3.5$, (c) $U=4.5$, and (d) $U=5.5$.
     }
    \label{fig:supercurrent and sde}
\end{figure}

Figure~\ref{fig:supercurrent and sde}(a) shows the magnetic field dependence of the superconducting diode efficiency $\eta$. 
The magnitude of $\eta$ increases with the magnetic field as usual, whereas it decreases with the onsite Coulomb interaction $U$.
Thus, the SDE is suppressed by the electron correlation effects.
In the following, we attribute this suppression to the strong coupling effects on the helical crossover phenomena.

The helical crossover, around which the properties of finite-momentum superconducting states drastically change~\cite{Dimitrova2003,Agterberg2007}, is expected to occur at relatively high fields around $h\sim0.08$ for $U=3.5$ and $h>0.1$ for $U=4.5$ and $5.5$, as discussed in the following.
Since our primary interest is the correlation-induced suppression of the SDE rather than the anomalous behavior associated with the helical crossover~\cite{Daido2022,Daido2022prb}, we restrict our analysis to the low-field regime $h<0.1$, 
which is sufficient for the present purpose.
Note that calculations for $U=5.5$ become difficult in the high-field regime  $h>0.06$ due to poor numerical convergence because the system is too close to the magnetic order, implying the field-induced AF order as observed in CeCoIn$_5$~\cite{Shishido2018}.

The intrinsic SDE in the Rashba-Zeeman model is characterized by the magnetic field of the helical crossover~\cite{Daido2022,Yuan2022,Daido2022prb,Ilic2022-it}.
Because of the asymmetric band structure, the helical superconductivity with a finite Cooper pair's momentum $p_0$ is stabilized~\cite{bauer2012non}, and $p_0$ drastically changes as the magnetic field increases through the helical crossover line~\cite{Dimitrova2003,Agterberg2007}.
The crossover field $h_{\rm cr}$ is roughly given by $h_{\rm cr} \sim \Delta$, where $\Delta$ is the superconducting gap~\cite{Daido2022,Yuan2022}.
Although this relation has been derived for $s$-wave superconductivity, it is expected to remain valid also for $d$-wave superconductivity~\cite{Daido2022prb}.
Hence, the magnetic field scaled by the crossover field is rewritten as
$
    h/h_{\rm cr}\sim h/\Delta 
    =\qty(T_{\rm c}h)/\qty(\Delta T_{\rm c}) 
    \equiv h/rT_{\rm c}
$,
where  
$r = \Delta/T_{\rm c}$ is the ratio between the superconducting gap and the critical temperature.

In quantum critical superconductors, the ratio $\Delta/T_{\rm c}$ becomes greater than the BCS value $r \sim 1.76$~\cite{YANASE20031} because the depairing effects arising from spin fluctuations are remarkably suppressed by a finite excitation gap below $T_{\rm c}$~\cite{Monthoux1994,Pao1994prl,Dahm1995prl,YANASE20031}, and the superconducting gap grows more than the weak-coupling BCS theory.
As the strong correlation effects grow with Coulomb interaction, 
the ratio $r = \Delta/T_{\rm c}$ increases with $U$~\cite{Pao1994prl}, although the transition temperature does not drastically change; $T_{\rm c}$ slightly increases from $\sim 0.011$ to $\sim 0.015$ with increasing $U$ from $3.5$ to $5.5$ in our model.
In contrast, we obtain the ratio $r=5$, $8$, and $11$ for $U=3.5$, $4.5$, and $5.5$, respectively,
where $\Delta$ is defined as the magnitude of the superconducting gap averaged for momentum $\sum_{\bm{k}} |\Delta(\bm{k})|/L^2$.
Thus, stronger electron correlations push the helical crossover line to the higher magnetic field region and reduce the scaled magnetic field $h/rT_{\rm c}$, which is consistent with the suppression of the SDE observed in the strongly correlated region.

The momentum of Cooper pairs $p_0$ in the helical superconducting state can be seen in Figs.~\ref{fig:supercurrent and sde}(b)-(d), which show the current-momentum ($j$-$q$) relations. 
Because the FLEX approximation is a conserving approximation, the supercurrent satisfies the thermodynamic relation $j(p)=\partial F(p)/\partial p$~\cite{supplemental}. 
Thus, the momentum $p_0$ that minimizes the free energy $F(p)$ is obtained from the relation $j(p_0)=0$.
Figure~\ref{fig:supercurrent and sde}(b) for $U=3.5$ reveals the drastic change of $p_0$, namely the helical crossover around $h\sim0.06$, which almost coincides with 
$rT_{\rm c} \sim 0.055$ evaluated by $r\simeq5$ and $T_{\rm c} \simeq 0.011$. 
We do not see the helical crossover in 
Figs.~\ref{fig:supercurrent and sde}(c) and (d), consistent with the evaluation $h_{\rm cr}\gtrsim0.1$ for $U=4.5$ and $5.5$.

\begin{figure}[tbp]
 \begin{center}
\includegraphics[height=5cm, width=8cm]{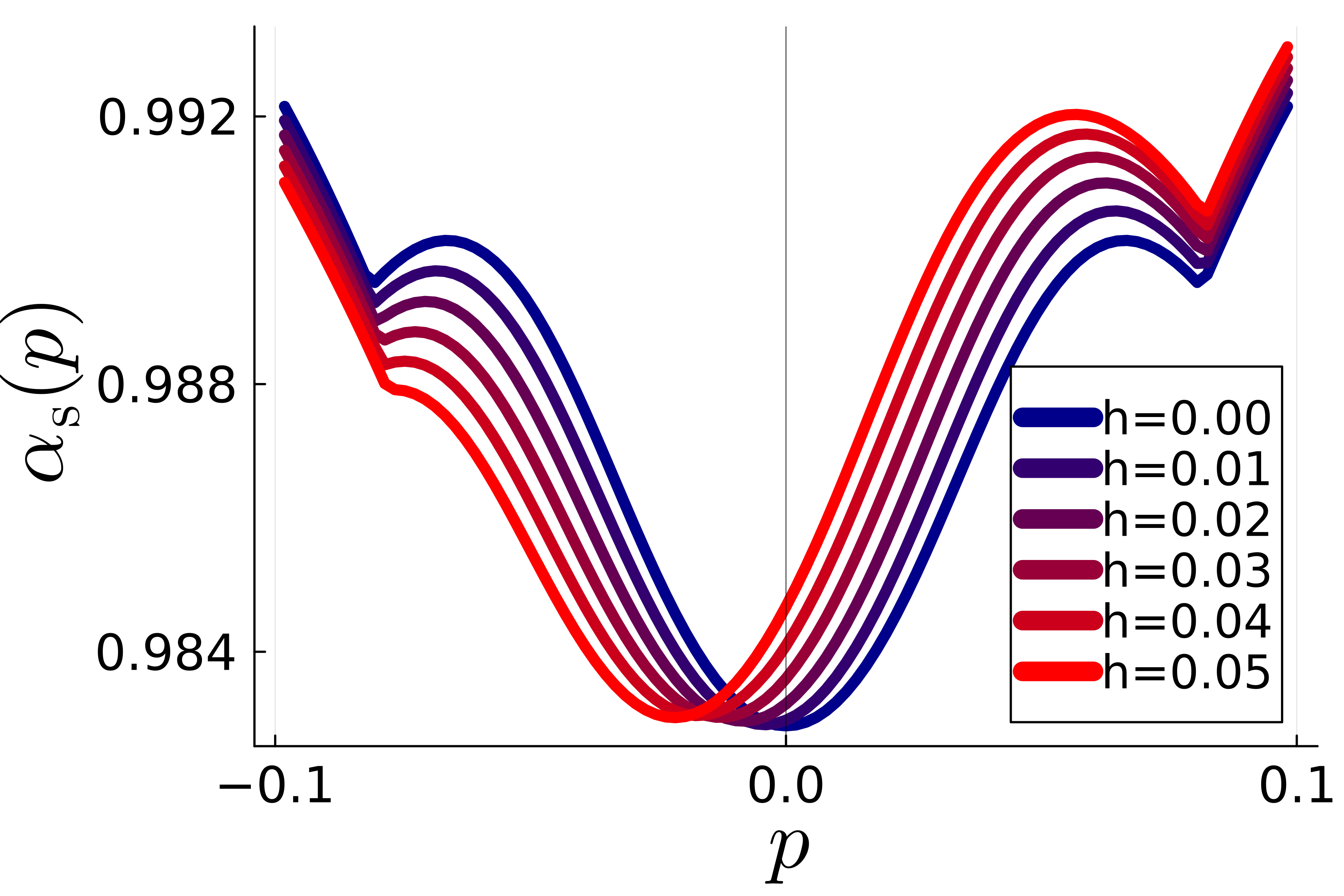}
  \end{center}
  \caption{
  Cooper pair's momentum dependence of the Stoner factor $\alpha_{\rm S}(p)$ for $U = 5.5$ and various magnetic fields.
  }
  \label{fig:stonerph}
\end{figure}

\textit{SDE by nonreciprocal magnetism.}---Here, we clarify the SDE tied to the magnetic order. Because the supercurrent nonreciprocally induces AF order, the SDE can originate from this nonreciprocal magnetism.
We also evaluate the superconducting diode efficiency derived from nonreciprocal magnetism and demonstrate the perfect diode efficiency $\eta=100$\%.

We first show the supercurrent-induced AF order and its nonreciprocity.
Figure~\ref{fig:stonerph} shows the Stoner factor as a function of the Cooper pair's momentum for $U=5.5$ and various magnetic fields. We see two significant features.
One notable feature is the asymmetry with respect to the Cooper pair's momentum, which is contrasted with almost reciprocal behaviors of the supercurrent $|j(p)| \sim |j(-p)|$
[compare Figs.~\ref{fig:supercurrent and sde}(d) and \ref{fig:stonerph}].
This nonreciprocity originates from the asymmetric deformation of the Fermi surface under supercurrent in the presence of the spin–orbit coupling and the Zeeman field, which leads to a directional dependence of the magnetic instability.
Another feature is that the Stoner factor approaches unity as the magnitude of the Cooper pair's momentum increases.
Since the Cooper pair's momentum is almost proportional to the supercurrent, this behavior implies the supercurrent-induced AF order.
Combining these two features, we recognize that the supercurrent can stabilize the AF order nonreciprocally, which is called {\it nonreciprocal magnetism} in the following.

We next discuss the SDE due to nonreciprocal magnetism.
In general, the magnetic order hardly coexists with superconductivity.
Even when coexistence occurs, as observed in a narrow window of underdoped cuprates~\cite{Ihara2025-iv},  
the critical current is significantly reduced and may fall below the applied current. 
We therefore assume that the AF order and superconductivity do not coexist in the current-carrying states.
Under this assumption, the supercurrent that induces the AF order sets an upper bound of the critical current, and thus nonreciprocal magnetism results in the SDE.
This mechanism is distinct from previously proposed SDE mechanisms based on depairing currents~\cite{Daido2022,Daido2022prb,He_2022,Yuan2022,Ilic2022-it}, vortex motion~\cite{Hou2023-if}, or nonequilibrium effects~\cite{Daido2025-uni}.

We now evaluate the superconducting diode efficiency by considering both nonreciprocal magnetism and depairing currents.
To do this, we have to determine the Stoner criterion for AF order and redefine the critical current. 
Although the formal Stoner criterion for the magnetic long-range order is $\alpha_{\rm S}=1$, within the FLEX approximation the magnetic instability is approached only asymptotically due to self-energy and mode-coupling effects~\cite{Moriya1990,Dahm1995}.
We therefore regard the Stoner factor as a measure of the proximity to an AF instability and introduce a phenomenological criterion $\alpha_{\rm c}=1-\delta_{\rm c}$, where $\delta_{\rm c}$ is a small positive parameter~\cite{Kino1999-pf}.
From Eqs.~\eqref{eq:stoner} and ~\eqref{eq:supercurrent}, the Stoner factor $\alpha_{\rm S}(j)$ is obtained as a function of the supercurrent $j$.
Then, the supercurrent that induces the AF order $j_{{\rm AF}\pm}$ is defined as
$
    \alpha_{\rm S}(j_{{\rm AF}\pm})
    =1 - \delta_{\rm c}
$,
based on the Stoner criterion discussed above.
Hence, the critical current is redefined as
$
    j_{\rm c+}=\min\qty[j_{\rm d+},j_{\rm AF+}]
$ and $
    j_{\rm c-}=\max\qty[j_{\rm d-},j_{\rm AF-}]
$.
The critical current is determined by the smaller of $j_{{\rm AF}\pm}$ and the depairing current $j_{{\rm d}\pm}$, and thus different types of SDE emerge depending on the parameters. 
Here, instead of changing physical parameters, we consider the SDE with varying the phenomenological parameter $\delta_{\rm c}$.

\begin{figure}[tbp]
    \centering
    \begin{tabular}{ll}
    { (a) $\delta_{\rm c} \lesssim0.008$, $|\eta|\sim3[\%]$ } &{ (b) $\delta_{\rm c} \sim0.01$, $|\eta|\sim35[\%]$} \\ 
     \includegraphics[keepaspectratio, scale=0.034]{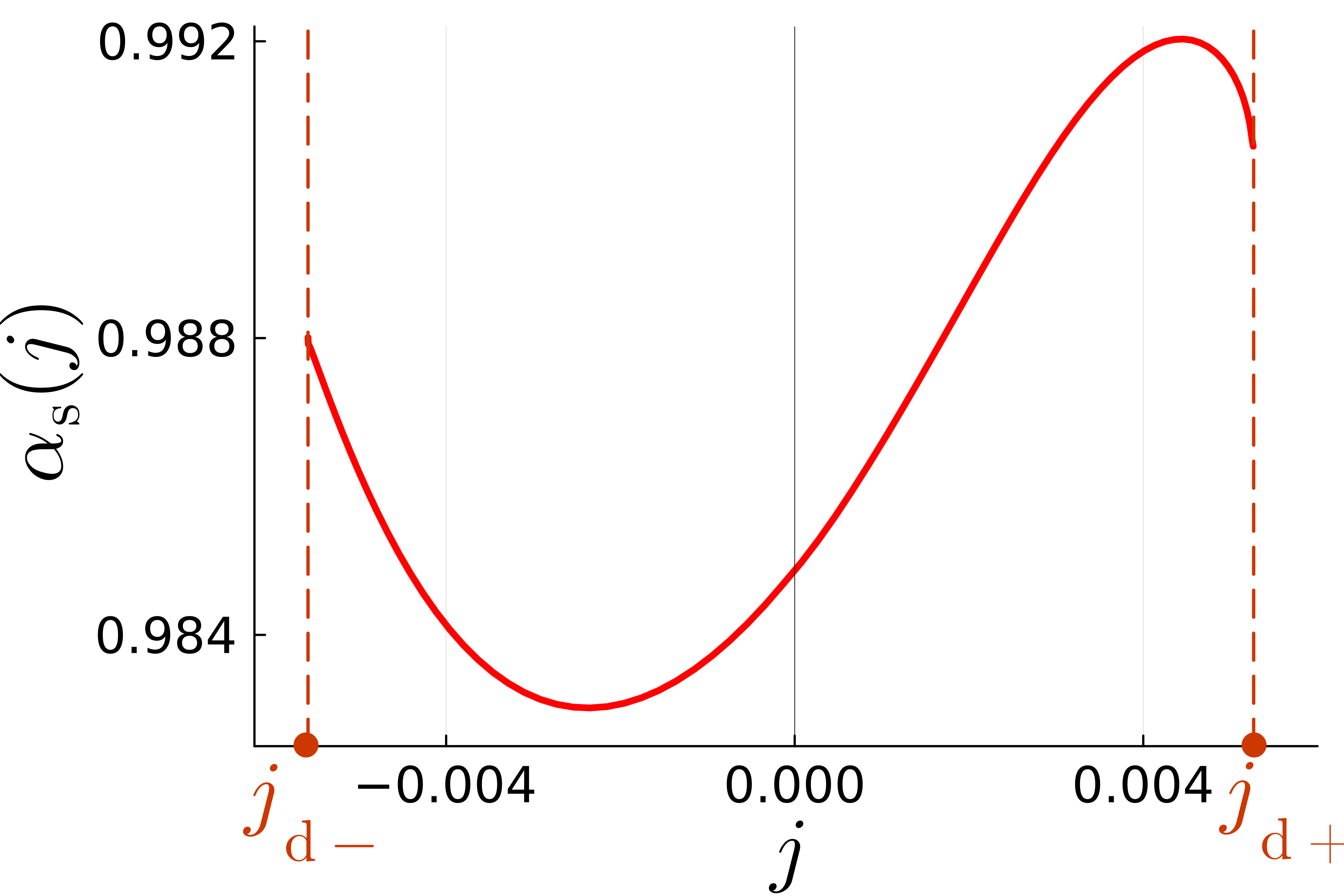} 
    &\includegraphics[keepaspectratio, scale=0.034]{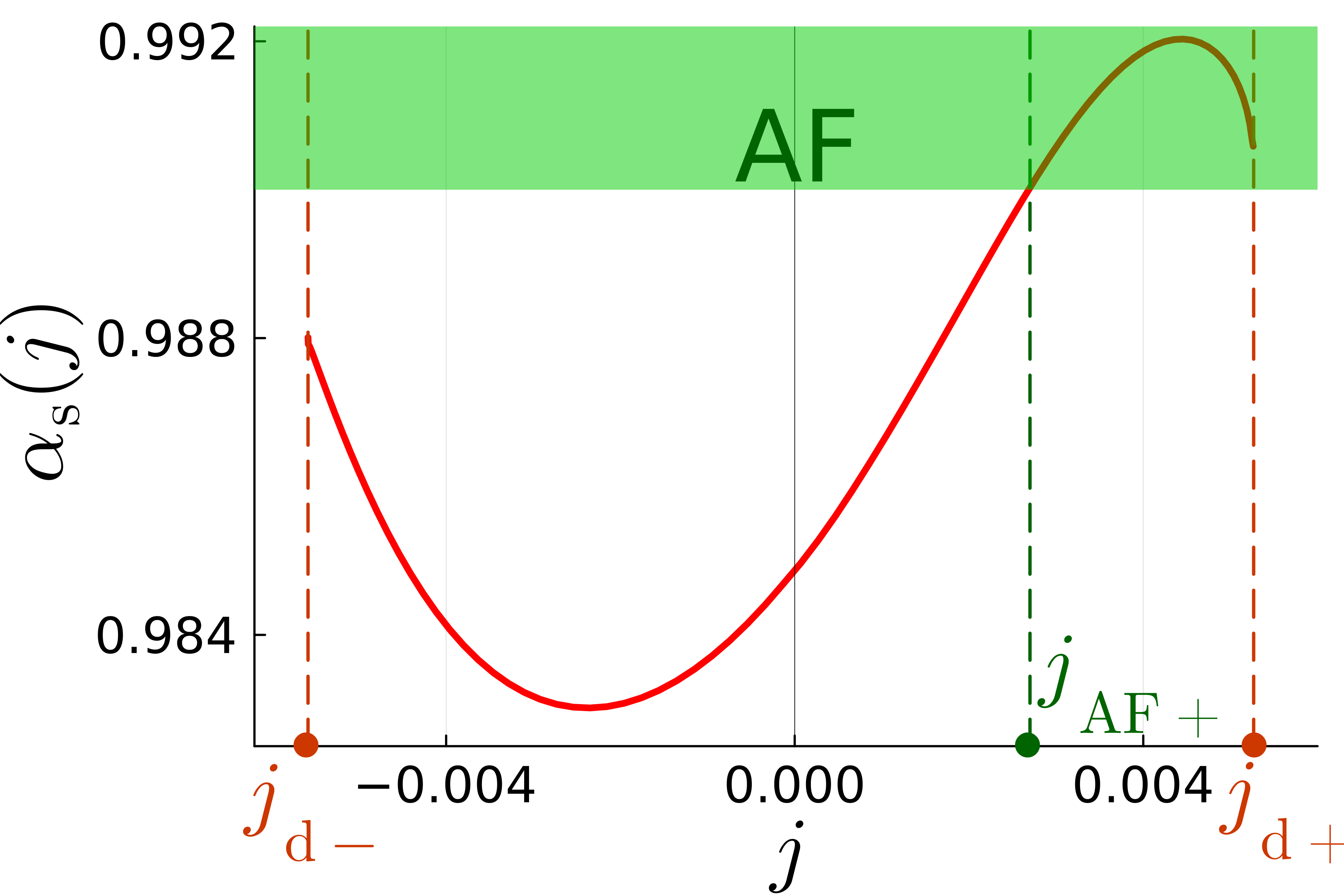} \\
    { (c) $\delta_{\rm c} \sim0.012$, $|\eta|\sim50[\%]$ } &{ (d) $\delta_{\rm c} \sim0.014$, $|\eta|\sim75[\%]$} \\ 
     \includegraphics[keepaspectratio, scale=0.034]{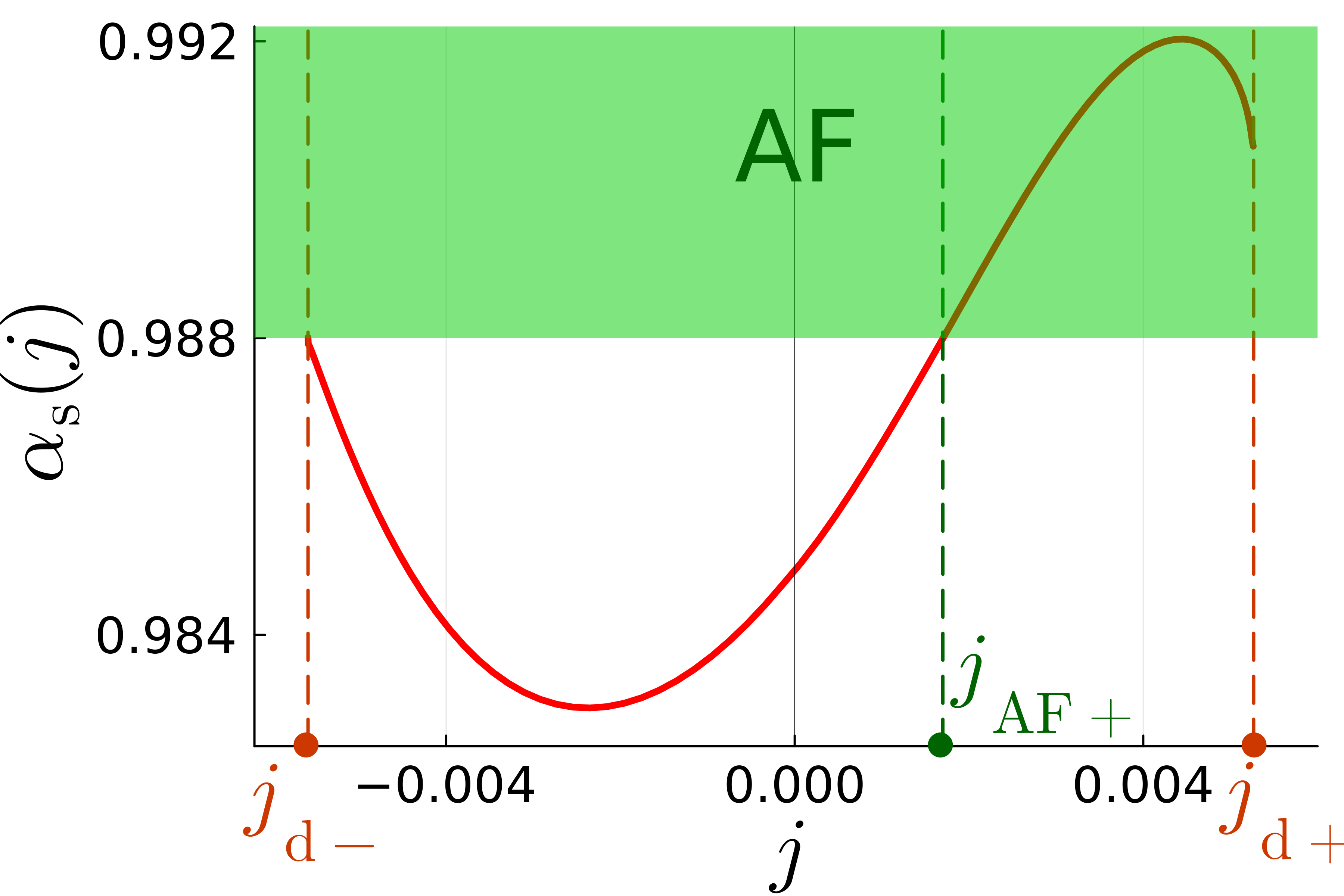} 
    &\includegraphics[keepaspectratio, scale=0.034]{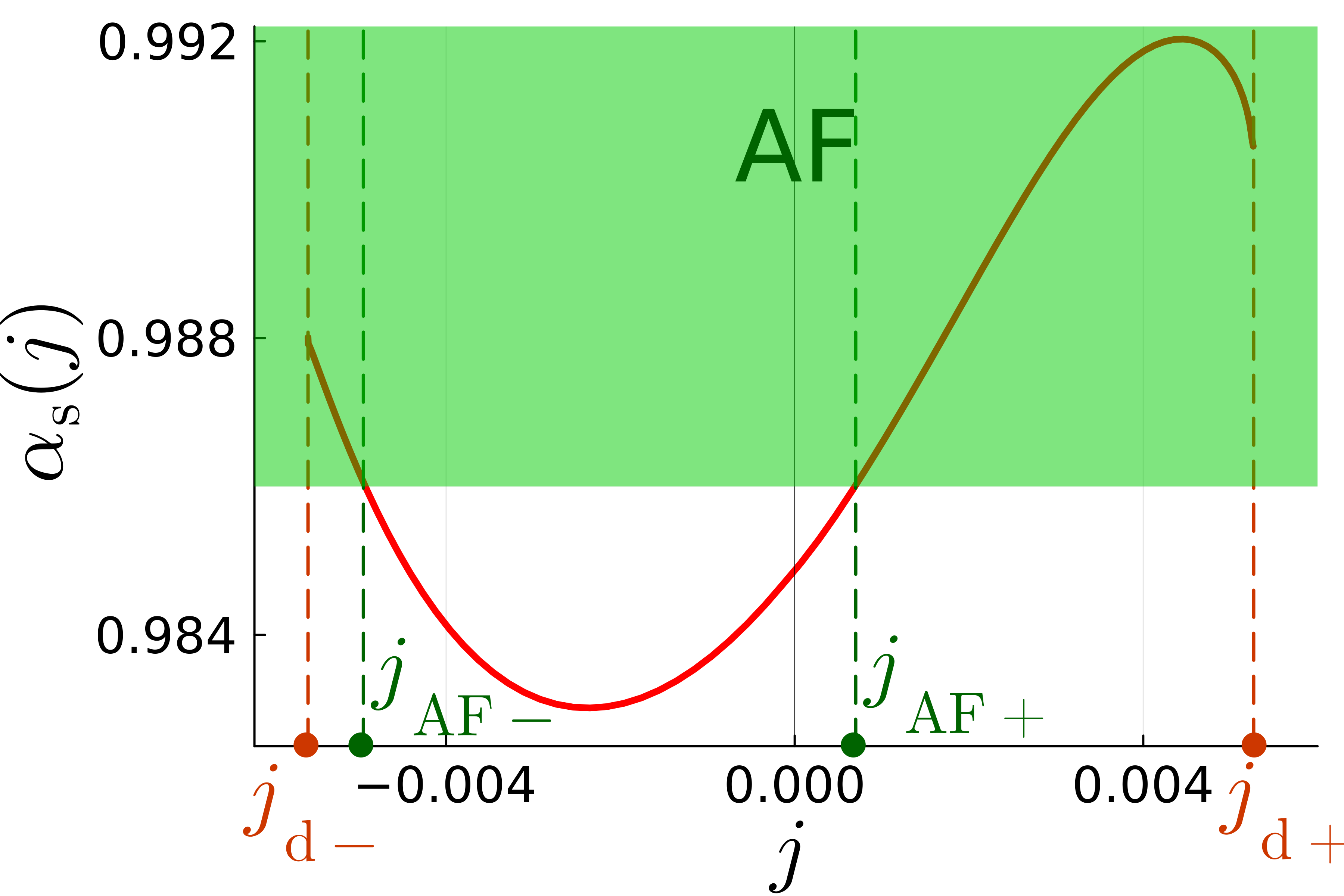}  \\
    { (e) $\delta_{\rm c} \sim0.015$, $|\eta|\sim100[\%]$} &{ (f) $\delta_{\rm c} \sim0.016$, $|\eta|\sim140[\%]$}\\ 
    \includegraphics[keepaspectratio, scale=0.034]{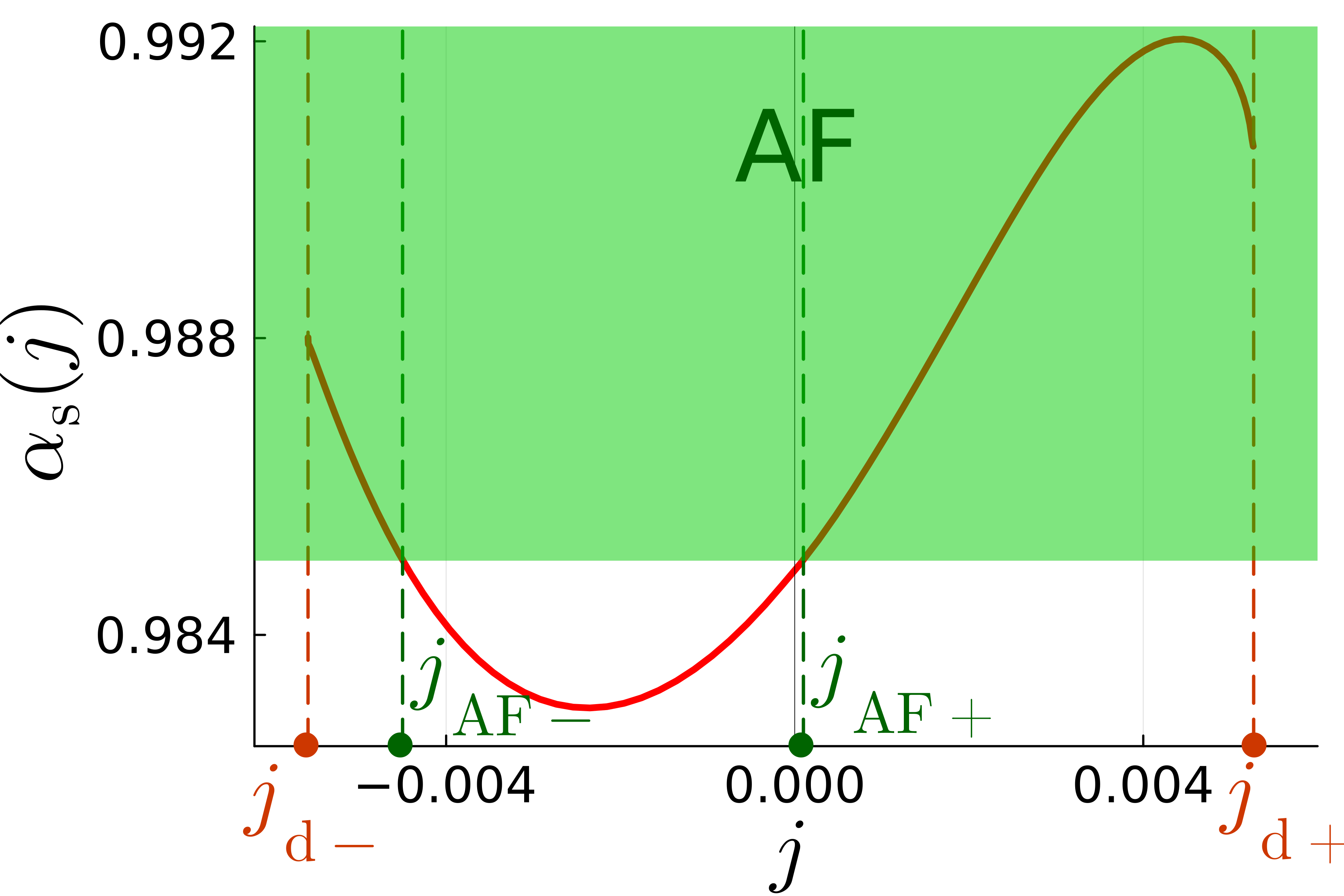} 
    &\includegraphics[keepaspectratio, scale=0.034]{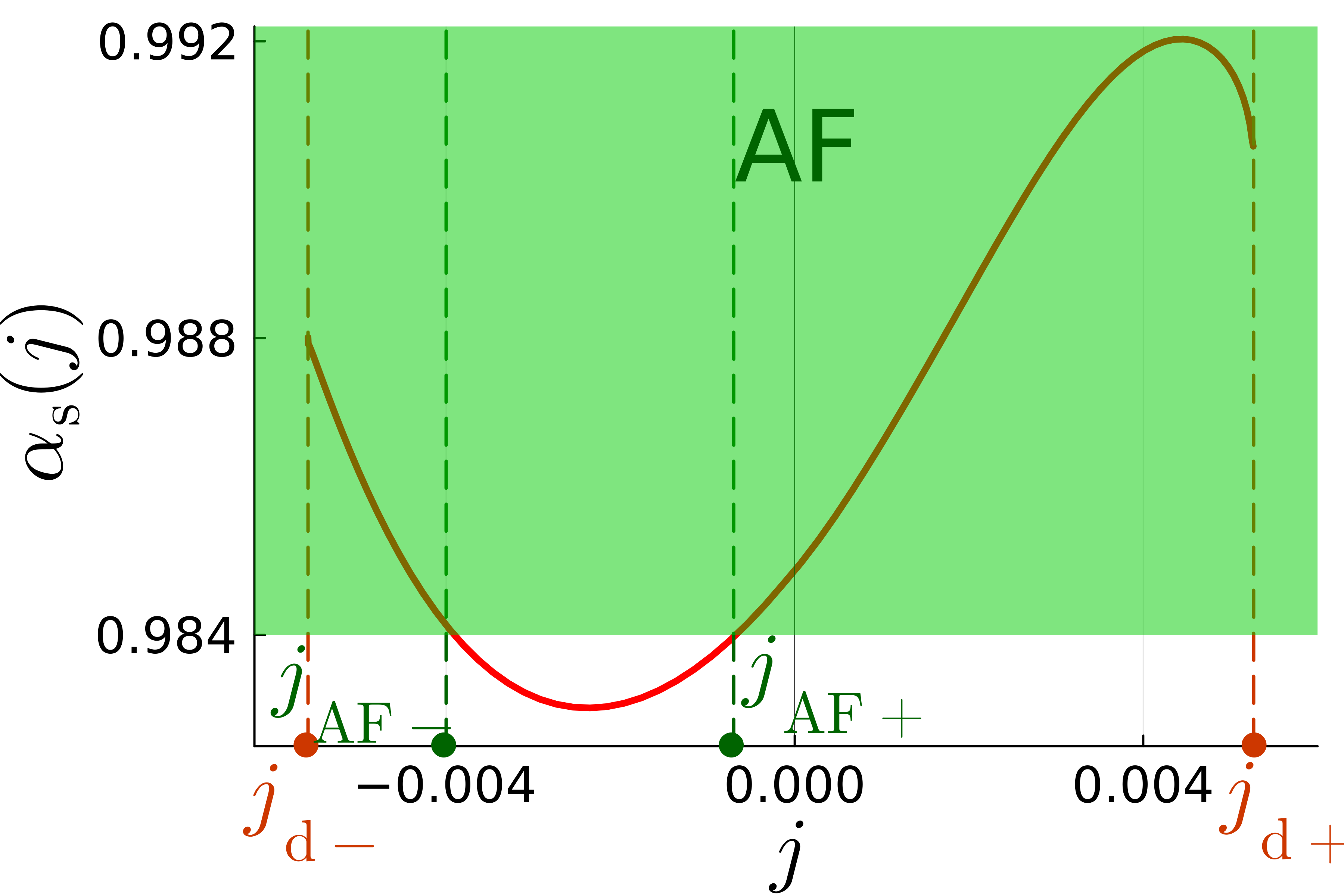}
    \end{tabular}
    \caption{
    Supercurrent dependence of the Stoner factor $\alpha_{\rm S}(j)$ for $U=5.5$ and $h=0.05$, where the parameter $\delta_{\rm c}$ is under (a) 0.008 and about (b) 0.01, (c) 0.012, (d) 0.014, (e) 0.015, and (f) 0.016.  
    The domain of $\alpha_{\rm S}(j)$ is the depairing current $j_{\rm d\pm}$, which is colored in orange.
    The region $\alpha_{\rm S}(j)>\alpha_{\rm c}=1-\delta_{\rm c}$ and the supercurrent that induces the AF order $j_{\rm AF\pm}$ are colored in light and dark green, respectively. 
    The diode efficiency is shown at the top of the figures.}
    \label{fig:sdeNM}
\end{figure}

In Fig.~\ref{fig:sdeNM}, we show the supercurrent dependence of the Stoner factor for $U=5.5$ and $h=0.05$.
Depending on the phenomenological parameter $\delta_{\rm c}$, the critical current is obtained as follows.
\begin{align}
    j_{\rm c+}&=
    \begin{cases}
        j_{\rm d+}, \quad &(\delta_{\rm c}\lesssim 0.008) \\
        j_{\rm AF+}>0, \quad &(0.008\lesssim\delta_{\rm c}\lesssim0.015) \\
        j_{\rm AF+}<0, \quad &(0.015\lesssim\delta_{\rm c}) 
    \end{cases}  
    \\
    j_{\rm c-}&=
    \begin{cases}
         j_{\rm d-}, \quad &(\delta_{\rm c}\lesssim 0.012) \\
        j_{\rm AF-}.  \quad &(0.012\lesssim\delta_{\rm c})
    \end{cases}
\end{align}
Thus, there are four regions in which the SDE arises from different mechanisms: (I) $\delta_{\rm c}\leq 0.008$, (II) $0.008 < \delta_{\rm c}\leq 0.012$, (III) $0.012 < \delta_{\rm c}\leq 0.015$, and (IV) $0.015 < \delta_{\rm c}$.
When we increase the Coulomb interaction $U$ or the carrier density, the Stoner factor increases, which is qualitatively equivalent to increasing $\delta_{\rm c}$.

The SDE in regions (I)-(IV) are illustrated in Fig.~\ref{fig:sdeNM}. 
In region (I), supercurrent-induced AF order does not occur  [Figure~\ref{fig:sdeNM}(a)]. Therefore, the critical current is determined by the depairing current,
$
    j_{\rm c\pm}=j_{\rm d\pm}
$,
and the diode efficiency is tiny around $3[\%]$ because the intrinsic SDE is suppressed by the electron correlation effects as discussed in the previous section.
In region (II) [Figs.~\ref{fig:sdeNM}(b) and (c)], the positive critical current is limited by the AF order,
$
    j_{\rm c+}=j_{\rm AF+}
$, 
while the negative one remains the depairing current,
$
    j_{\rm c-}=j_{\rm d-}
$.
As a result, the diode efficiency is significantly improved and reaches $50[\%]$ in Fig.~\ref{fig:sdeNM}(c). 
In region (III) [Figs.~\ref{fig:sdeNM}(d) and (e)], both critical currents
are determined by the AF order,
$
    j_{\rm c\pm}=j_{\rm AF\pm} 
$. 
As the system approaches the AF order, 
$j_{\rm AF+}$ approaches zero, but $j_{\rm AF-}$ remains finite, leading to the perfect diode efficiency $100[\%]$ [Fig.~\ref{fig:sdeNM}(e)]. 
Thus, the perfect SDE can be realized through nonreciprocal magnetism.

A further striking phenomenon emerges in region (IV).
There, the critical current defined for the positive direction becomes negative, so the superconducting state is stable only under the negative current [Fig.~\ref{fig:sdeNM}(f)]. This phenomenon corresponds to the current-induced zero-resistance state~\cite{daidoCIZRS}.  
At zero current ($j=0$), the AF state is stable. However, the superconducting state appears in the finite-current region $j_{{\rm AF}-}<j<j_{{\rm AF}+}<0$. 
In this case, the current induces superconductivity rather than magnetism. 

\textit{Discussion.}---The supercurrent enhances low-energy quasiparticles in nodal superconductors and suppresses the superconducting gap due to an increase in kinetic energy, thereby promoting magnetic instability near an AF quantum critical point~\cite{Nakamura2025}.
This principle is broadly applicable to strongly correlated superconductors near magnetic instabilities.
It may also be viewed as a coupling between supercurrent and symmetry-breaking order parameters~\cite{Banerjee2024} in SCES. 
Whereas Ref.~\cite{Banerjee2024} has studied the modulation of preexisting order by a supercurrent, our results demonstrate that a supercurrent can dynamically induce AF order.

Essential ingredients for nonreciprocal magnetism are nodal superconductivity, proximity to the AF instability, and broken inversion and time-reversal symmetries. 
Among candidate materials, high-$T_{\rm c}$ cuprate superconductors are particularly suitable because they naturally host nodal $d$-wave superconductivity and lie close to AF quantum criticality.
Broken inversion and time-reversal symmetries may be realized, for example, in cuprate-manganite heterostructures through interfacial Rashba coupling and exchange fields~\cite{GOLDMAN199969,Chakhalian2006-tg,Nemes2008},
or in cuprates with loop-current order~\cite{Varma2006,Maruyama2015,Watanabe2021LC,varma2025antisymmetricchiralcurrentszero,Tazai2025},
which may be related to the recently observed field-free SDE~\cite{Qi2025} in cuprates.

Our findings highlight not only the impact of strong correlations and quantum criticality on the SDE but also the role of the supercurrent as a tuning knob. Our results establish a design principle for realizing large and even perfect SDE in strongly correlated superconductors via nonreciprocal magnetism.
More broadly, they extend the scope of the theoretical study of SDE to SCES.

\begin{acknowledgments} 
This work was supported by JSPS KAKENHI (Grant Numbers JP22H01181, JP22H04933,
JP23K17353, JP23K22452, JP24K21530, JP24H00007, JP24KJ1475, JP25H01249, JP26H02016).
\end{acknowledgments}

\bibliography{ref.bib}

\clearpage

\renewcommand{\bibnumfmt}[1]{[S#1]}
\renewcommand{\citenumfont}[1]{S#1}
\renewcommand{\thesection}{S\arabic{section}}
\renewcommand{\theequation}{S\arabic{equation}}
\setcounter{equation}{0}
\renewcommand{\thefigure}{S\arabic{figure}}
\setcounter{figure}{0}
\renewcommand{\thetable}{S\arabic{table}}
\setcounter{table}{0}
\makeatletter
\c@secnumdepth = 2
\makeatother

\onecolumngrid

\begin{center}
 {\large \textmd{Supplemental Material:} \\[0.3em]
 {\bfseries Superconducting diode effect in correlated electron systems by nonreciprocal magnetism}}
\end{center}

\setcounter{page}{1}

\twocolumngrid

\section{Dyson-Gor'kov equation and fluectuation exchange approximation}
\label{appendix:flex}
Here we formulate the Dyson-Gor'kov equation and the fluctuation exchange (FLEX) approximation for the Rashba-Zeeman-Hubbard model.
The formulation is mainly based on Ref.~\cite{Kita2011-at}.

The Dyson-Gor'kov equation consists of bare Green functions $\hat{\mathcal{G}}_0$, dressed Green functions $\hat{\mathcal{G}}$, and self-energies $\hat{\varsigma}$, which are $4\times4$ matrices 
spanned by spin and Nambu space in the present case. 
These matrices are introduced by
\begin{align}
    \hat{\mathcal{G}}_0(k,\bm{p})=\mqty(i\omega_n \bm{I}_2-\ul{\xi}(\bm{k}+\bm{p}) & \bm{0} \\ \bm{0} & i\omega_n\bm{I}_2+\ul{\xi}^T(-\bm{k}+\bm{p}))^{-1} , \label{eq:greenzero}
\end{align}
\begin{align}
    \hat{\mathcal{G}}(k,\bm{p})=\mqty( \ul{G}(k,\bm{p}) & \ul{F}(k,\bm{p}) \\ \ul{\tilde{F}}(k,\bm{p}) & \ul{\tilde{G}}(k,\bm{p})) ,
\end{align}
and
\begin{align}
    \hat{\varsigma}(k,\bm{p})=\mqty( \ul{\Sigma}(k,\bm{p}) & \ul{\Delta}(k,\bm{p}) \\ \ul{\tilde{\Delta}}(k,\bm{p}) & \ul{\tilde{\Sigma}}(k,\bm{p})),
\end{align}
where $k=(\bm{k},i\omega_n)$ with the fermionic Matsubara frequency $i\omega_n=(2n+1)\pi T$, and $\ul{G}$, $\ul{F}$, $\ul{\Sigma}$, and $\ul{\Delta}$, respectively, represent $2\times2$ matrices of normal Green function, anomalous Green function, normal self-energy, and anomalous self-energy.
The off-diagonal elements of $\ul{G}$ and $\ul{\Sigma}$ and the diagonal elements of $\ul{F}$ and $\ul{\Delta}$ are non-zero because of the Rashba spin-orbit coupling that causes parity-mixing in superconducting order parameter~\cite{bauer2012non}.
The Dyson-Gor'kov equation is expressed as
\begin{align}
    \hat{\mathcal{G}}(k,\bm{p})=\qty{\hat{\mathcal{G}}_0(k,\bm{p})^{-1}-\hat{\varsigma}(k,\bm{p})}^{-1}. \label{eq:DG eq}
\end{align}

From the Dyson-Gor'kov equation [Eq.~\eqref{eq:DG eq}], the Green functions are determined by the self-energies.
In the Luttinger-Ward formalism, the self-energies are obtained by the generating function 
$\Phi[\hat{\mathcal{G}}]$ as follows~\cite{Kita2011-at,Luttinger1960,YANASE20031},
\begin{align}
    \varsigma_{\mu\nu}(k,\bm{p})=2\frac{\delta\Phi[\hat{\mathcal{G}}]}{\delta \mathcal{G}_{\nu\mu}(k,\bm{p})}. \label{eq:variation eq}
\end{align}
The factor $2$ is derived here by individually handling the matrices with and without the tilde.
To formulate the Luttinger-Ward generating function, some approximation is necessary.
Then, we adopt the FLEX approximation for the following two reasons.
First, the FLEX approximation incorporates critical spin fluctuations, which in turn captures the essence of the AF quantum critical $d$-wave superconductivity~\cite{Moriya01072000,YANASE20031}.
Second, the FLEX approximation is a conserving approximation~\cite{Baym1961} that satisfies thermodynamic relations, and thus we can uniquely evaluate thermodynamic quantities such as the supercurrent, which is necessary to investigate the SDE, from a thermodynamic potential (see Sec.~\ref{appendix:supercurrent}).

To formulate the FLEX approximation, we introduce $8\times8$ matrices of bare interaction $\check{U}$ and bare spin susceptibility $\check{\chi}^0(q,\bm{p})$ as follows. 
\begin{align}
    \check{U}=\frac{U}{2}\mqty(\hat{U}_a & \hat{U}_b \\ \hat{U}_b & \hat{U}_a),
\end{align}
\begin{align}
    \check{\chi}^0(q,\bm{p})=\mqty(\hat{\chi}^0_{GG}(q,\bm{p}) & \hat{\chi}^0_{F\tilde{F}}(q,\bm{p}) \\ \hat{\chi}^0_{\tilde{F}F}(q,\bm{p}) & \hat{\chi}^0_{\tilde{G}\tilde{G}}(q,\bm{p})),
\end{align}
where $q=(\bm{q},i\Omega_n)$ with the bosonic Matsubara frequency $i\Omega_n=2n\pi T$, 
\begin{align}
    \hat{U}_a=\mqty(0&0&0&1 \\ 0&-1&0&0 \\ 0&0&-1&0 \\ 1&0&0&0), \> \hat{U}_b=\mqty(0&0&0&-1\\ 0&0&1&0 \\ 0&1&0&0 \\ -1&0&0&0),
\end{align}
and
\begin{align}
    \qty[\hat{\chi}^0_{AB}(q,\bm{p})]_{\alpha\beta,\gamma\delta}=-\sum_k A_{\alpha\gamma}(k+q,\bm{p})B_{\delta\beta}(k,\bm{p}).
\end{align}
Here, $\sum_k$ means $T/L^2\sum_{\bm{k},i\omega_n}$.
The Luttinger-Ward generating function under the FLEX approximation is represented as~\cite{Kita2011-at}
\begin{align}
    \Phi[\hat{\mathcal{G}}]=&\frac{1}{2}\sum_q \left\{\log\det\qty[\bm{I}_8+\check{U}\check{\chi}^0(q,\bm{p)}] -\mathrm{Tr}\qty[\check{U}\check{\chi}^0(q,\bm{p})] \right\} \nonumber \\
    &+ \frac{U^2}{4}\sum_{q,\alpha} \left\{ \qty[\hat{\chi}^0_{GG}(q,\bm{p})-\hat{\chi}^0_{F\tilde{F}}(q,\bm{p})]_{\uparrow\uparrow,\alpha\alpha}
    \right. \nonumber  \\
    & \left. \qquad\quad \times \qty[\hat{\chi}^0_{GG}(q,\bm{p})-\hat{\chi}^0_{F\tilde{F}}(q,\bm{p})]_{\downarrow\downarrow,-\alpha-\alpha}   \right. \nonumber \\
    & \left. \qquad\quad + \qty[\hat{\chi}^0_{\tilde{G}\tilde{G}}(q,\bm{p})-\hat{\chi}^0_{\tilde{F}F}(q,\bm{p})]_{\uparrow\uparrow,\alpha\alpha}   \right. \nonumber \\
    & \left. \qquad\quad \times \qty[\hat{\chi}^0_{\tilde{G}\tilde{G}}(q,\bm{p})-\hat{\chi}^0_{\tilde{F}F}(q,\bm{p})]_{\downarrow\downarrow,-\alpha-\alpha}    \right\}, \label{eq:LW funtion}
\end{align}
where $\sum_q$ means $T/L^2\sum_{\bm{q},i\Omega_n}$, and the term proportional to $U^2$ is introduced to avoid double counting of the second order ladder and bubble skeleton diagrams.
From Eqs.~\eqref{eq:DG eq}, \eqref{eq:variation eq}, and \eqref{eq:LW funtion}, the Green functions and self-energies are self-consistently determined.

Finally, we comment on parameters used in numerical calculations. 
The temperature is set to $T=0.005<T_{\rm c} \sim0.015$ so as to be much lower than the superconducting critical temperature $T_{\rm c}$.
The system size is set to $L^2=64\times64$.
For Matsubara frequencies, sparse-ir is adopted for efficient numerical calculations~\cite{Shinaoka2017,LiJia2020,Shinaoka2022}.

\section{Supercurrent and free energy}
\label{appendix:supercurrent}

We show that the supercurrent is expressed as the derivative of free energy with respect to Cooper pair's momentum.

The thermodynamic potential $\Omega(T,\mu,\vb{p})$ is formulated based on the Luttinger-Ward formalism and given by
\begin{align}
    \Omega(T,\mu,\bm{p})=\Omega_0(T,\mu,\bm{p})+\Omega_F(T,\mu,\bm{p})+\Phi[\hat{\mathcal{G}}],
\end{align}
where 
\begin{align}
    \Omega_0(T,\mu,\bm{p})=-\frac{1}{2}\sum_k\log\mathrm{det}\qty[\hat{\mathcal{G}}_0^{-1}] +\mathrm{const}.,
\end{align}
\begin{align}
    \Omega_F(T,\mu,\bm{p})=-\frac{1}{2}\sum_k\left\{\log\frac{\det \hat{\mathcal{G}}(k,\bm{p})^{-1}}{\det \hat{\mathcal{G}}_0(k,\bm{p})^{-1}} \right. \nonumber \\
    \left. +\Tr\qty[\hat{\mathcal{G}}(k,\bm{p})\hat{\zeta}(k,\bm{p})]\right\} ,
\end{align}
and $\Phi[\hat{\mathcal{G}}]$ is given by Eq.~\eqref{eq:LW funtion}.
The free energy is obtained by
\begin{align}
    F(T,n,\bm{p})=\Omega(T,\mu,\bm{p})+\mu n.
\end{align}

We calculate the derivative of the free energy with respect to the momentum of Cooper pairs. 
The derivative of $\Omega_0(T,\mu,\bm{p})+\Omega_F(T,\mu,\bm{p})$ is given by
\begin{align}
    &\pdv{p_x}\qty{\Omega_0(T,\mu,\bm{p})+\Omega_F(T,\mu,\bm{p})} \nonumber \\
    &=-\frac{1}{2}\pdv{p_x}\sum_k\qty{\log\det\hat{\mathcal{G}}(k,\bm{p})^{-1}+\Tr\qty[\hat{\mathcal{G}}(k,\bm{p})\hat{\zeta}(k,\bm{p})]} \nonumber \\
    &=-\frac{1}{2}\sum_k\left\{\Tr\qty[\pdv{\hat{\mathcal{G}}(k,\bm{p})^{-1}}{p_x}\hat{\mathcal{G}}(k,\bm{p})] \right. \nonumber \\
    &\left. +\Tr\qty[\pdv{\hat{\mathcal{G}}(k,\bm{p})}{p_x}\hat{\zeta}(k,\bm{p})]
    +\Tr\qty[\hat{\mathcal{G}}(k,\bm{p})\pdv{\hat{\zeta}(k,\bm{p})}{p_x}]
    \right\} . \label{eq:pdv omegazf}
\end{align}
From Eq.~\eqref{eq:variation eq}, the derivative of $\Phi[\hat{\mathcal{G}}]$ is obtained as
\begin{align}
    \pdv{\Phi[\hat{\mathcal{G}}]}{p_x}&=\sum_k\frac{\delta\Phi[\hat{\mathcal{G}}]}{\delta \mathcal{G}_{\mu\nu}(k,\bm{p})}
    \pdv{\mathcal{G}_{\mu\nu}(k,\bm{p})}{p_x} \nonumber \\
    &=\frac{1}{2}\sum_k\Tr\qty[\pdv{\hat{\mathcal{G}}(k,\bm{p})}{p_x}\hat{\zeta}(k,\bm{p})] . \label{eq:pdv LW}
\end{align}
From Eqs.~\eqref{eq:pdv omegazf}, ~\eqref{eq:pdv LW}, ~\eqref{eq:DG eq}, and ~\eqref{eq:verocity}, the derivative of the thermodynamic potential is given by
\begin{align}
 \pdv{\Omega(T,\mu,\bm{p})}{p_x}  
    &=-\frac{1}{2}\sum_k\left\{\Tr\qty[\pdv{\hat{\mathcal{G}}(k,\bm{p})^{-1}}{p_x}\hat{\mathcal{G}}(k,\bm{p})] \right. \nonumber \\
    &\left. \qquad+\Tr\qty[\hat{\mathcal{G}}(k,\bm{p})\pdv{\hat{\zeta}(k,\bm{p})}{p_x}]
    \right\}  \nonumber \\   
    &=-\frac{1}{2}\sum_k\Tr\qty[\pdv{\hat{\mathcal{G}}_0(k,\bm{p})^{-1}}{p_x}\hat{\mathcal{G}}(k,\bm{p})] \nonumber \\
    &=\frac{1}{2}\sum_k\Tr\qty[\pdv{\hat{\Xi}(\bm{k},\bm{p})}{p_x}\hat{\mathcal{G}}(k,\bm{p})] .
    \label{eq:pdv omega}
\end{align}
The derivative of $\hat{\Xi}$ is transformed into
\begin{align}
    \pdv{\hat{\Xi}(\bm{k},\bm{p})}{p_x}&=\pdv{\hat{\Xi}(\bm{k},\bm{p})}{k_x}-\pdv{\mu}{p_x}\hat{M},
\end{align}
where
\begin{align}
    \hat{M}=\mqty(1&0&0&0 \\ 0&1&0&0 \\ 0&0&-1&0 \\ 0&0&0&-1).
\end{align}
Then, Eq.~\eqref{eq:pdv omega} is transformed into
\begin{align}
    \pdv{\Omega(T,\mu,\bm{p})}{p_x}&=\frac{1}{2}\sum_k\Tr\qty[\pdv{\hat{\Xi}(\bm{k},\bm{p})}{k_x}\hat{\mathcal{G}}(k,\bm{p})] \nonumber \\
    &\quad -\pdv{\mu}{p_x}\sum_k\Tr\qty[\frac{\hat{M}}{2}\hat{\mathcal{G}}(k,\bm{p})] \nonumber \\
    &=\frac{1}{2}\sum_k\Tr\qty[\pdv{\hat{\Xi}(\bm{k},\bm{p})}{k_x}\hat{\mathcal{G}}(k,\bm{p})]-\pdv{\mu}{p_x}n .
\end{align}
Finally, the derivative of the free energy is shown to be equivalent to the current, 
\begin{align}
    \pdv{F(T,n,\bm{p})}{p_x}&=\frac{1}{2}\sum_k\Tr\qty[\pdv{\hat{\Xi}(\bm{k},\bm{p})}{k_x}\hat{\mathcal{G}}(k,\bm{p})] \nonumber \\
    &=j({\bm p}) .
\end{align}
Therefore, when the free energy $F(\bm{p})$ takes the minimum value at $\bm{p}=\bm{p}_0$, the relation $j(\bm{p}=\bm{p}_0)=0$ is satisfied.
In other words, the Cooper pair's momentum that satisfies $j(\bm{p})=0$ is realized in the stable superconducting state, and $\bm{p}_0$ is called the stable momentum of Cooper pairs.

\end{document}